\documentclass[epjCONF,columns]{svjour} 
\usepackage{graphics}
\usepackage[varg]{txfonts} 
\usepackage[latin1]{inputenc}
\session-title{NSRT12}

\def\q{\mbox{\boldmath $q$}}
\def\mcv{\mbox{$\mathcal{V}$}}
\def\mch{\mbox{$\mathcal{H}$}}
\newcommand{\diff}{{\rm\,d}}                    

\begin{document}
\title{Relativistic models for quasielastic electron and neutrino-nucleus
scattering}
\author{Carlotta Giusti\inst{1}\fnmsep\thanks{\email{Carlotta.Giusti@pv.infn.it.}}
 \and Andrea Meucci\inst{1}}
%
\institute{Dipartimento di Fisica, 
Universit\`{a} degli Studi di Pavia and \\
INFN, Sezione di Pavia, via Bassi 6 I-27100 Pavia, Italy}
\abstract{Relativistic models developed within the framework of the impulse 
approximation for quasielastic (QE) electron scattering and successfully 
tested in comparison with electron-scattering data have been extended to 
neutrino-nucleus scattering.
Different descriptions of final-state interactions (FSI) in the 
inclusive scattering are compared. In the relativistic Green's function (RGF) 
model FSI are described consistently with the 
exclusive scattering using a complex optical potential. In the relativistic 
mean field  (RMF) model FSI are described by the same RMF potential which 
gives the bound states. The results of the models are compared for electron 
and neutrino scattering and, for neutrino scattering, with the recently 
measured charged-current QE (CCQE) MiniBooNE 
cross sections. 
} 
\maketitle
\section{Introduction}
\label{intro}
The analysis of the recent neutrino-nucleus scattering cross sections 
measured by the MiniBooNE Collaboration~\cite{miniboone,miniboonenc} requires 
reliable theoretical models where all nuclear effects are well under control. 
The energy region explored requires a relativistic description of the process, 
where not only relativistic kinematics is considered but also nuclear dynamics 
and current operators are described within a relativistic framework. 
Within the QE kinematics domain, the treatment of the FSI between the ejected 
nucleon and the residual nucleus is an essential ingredient for the comparison 
with data.  

Relativistic models for QE electron and neutrino-nucleus scattering, with different
descriptions of FSI, are presented and compared in this contribution.

In spite of many similar aspects, electron and neutrino-nucleus
scattering are two different processes and it is 
not guaranteed that a model able to describe electron scattering data will be 
able to describe neutrino scattering data with the same accuracy. 
The large amount of data available for 
electron scattering and all the theoretical and experimental work done 
over several decades, which provided a wealth of detailed information on 
nuclear structure and dynamics~\cite{rep,book}, make electron 
scattering the best available guide to determine the predictive power of a 
nuclear model. A good description of electron scattering data can therefore be
considered a necessary prerequisite for any model aimed to describe 
neutrino-nucleus cross sections.

In the QE region the nuclear response is dominated by the mechanism of 
one-nucleon knockout, where the probe interacts with a quasifree 
nucleon which is emitted from the nucleus with a direct one-step mechanism and 
the remaining nucleons are spectators.
In the exclusive $(e,e^{\prime}p)$  reaction the outgoing nucleon is detected in 
coincidence with the scattered electron, the residual nucleus is left in a 
specific discrete eigenstate, and the final state is completely 
specified. In the inclusive $(e,e^{\prime})$ scattering only the scattered
electron is detected, the final nuclear state is not determined, and the cross 
section includes all the available final nuclear states. 
For an incident neutrino, neutral-current (NC) and charged-current
(CC) scattering can be considered.  
In NC scattering only the emitted nucleon can be detected and the process is
inclusive in the lepton sector (the final neutrino is not detected) and
semi-inclusive in the hadron sector (the emitted nucleon is detected but the
final nuclear state is not completely determined).
In CC scattering the inclusive scattering where only the final lepton is 
detected can be treated with the same models used for the inclusive 
$(e,e^{\prime})$ reaction. 

For all these processes the cross section is obtained  in the one-boson 
exchange approximation from the contraction between the lepton tensor, which,
under the assumption of the plane-wave approximation for the initial and the 
final lepton wave functions, depends  only on the lepton kinematics, and the 
hadron tensor, whose components are given by bilinear products of the matrix 
elements of the nuclear current between the initial and final nuclear states.
Different but consistent models to calculate the hadron tensor in the exclusive
and in the inclusive scattering have been developed and are outlined in the
following sections.

\section{The exclusive $(e,e^{\prime}p)$ reaction}
\label{sec:1}

The theoretical framework for the description of the exclusive 
$(e,e^{\prime}p)$ reaction is the distorted-wave impulse approximation (DWIA). 
The model is based on the following assumptions~\cite{rep,book}: 
\newline
i) An exclusive process is considered, where the residual nucleus is left in a
discrete eigenstate $n$ of its Hamiltonian.
\newline
ii) The final nuclear state is projected onto the channel subspace spanned by 
the vectors corresponding to a nucleon, at a given position, and the residual 
nucleus in the state $n$. 
\newline
iii) The (one-body) nuclear-current operator does not connect different channel 
subspaces and also the initial state is projected onto the selected channel 
subspace. This is the assumption of the direct-knockout mechanism 
and of the IA.


The matrix elements of the nuclear current are then obtained in a
one-body representation as

\begin{equation}
\lambda_n^{1/2} \langle\chi^{(-)}\mid  j^{\mu}(\q) \mid \varphi_n \rangle  \ ,
\label{eq.dko}
\end{equation}  
where $j^{\mu}$  the one-body nuclear current, $\chi^{(-)}$ is the 
single-particle (s.p.) scattering state of the emitted nucleon, $\varphi_n$ the 
overlap between the ground state of the target and the final state $n$, 
i.e., a s.p. bound state, and the spectroscopic factor $\lambda_n$ is the norm
of the overlap function, that gives the probability of removing a nucleon from 
the target leaving the residual nucleus in the state $n$. 
The s.p. bound and scattering states are 
eigenfunctions of a non Hermitian energy-dependent Feshbach-type optical 
potential and of its Hermitian conjugate at different energies. 
In standard DWIA calculations phenomenological ingredients are usually employed: 
the scattering states are eigenfunctions of a phenomenological optical potential 
determined through a fit to elastic nucleon-nucleus scattering data and the 
s.p. bound states are usually obtained from mean-field potentials. 

The model can be formulated in a similar way within nonrelativistic DWIA 
\cite{bof82} and relativistic RDWIA frameworks \cite{meucci01}. 
In RDWIA, calculations 
are performed with a relativistic nuclear-current operator and four-vector 
relativistic wave functions for the s.p. bound and scattering states. 
Both the DWIA and the RDWIA have been quite successful in describing 
$(e,e^{\prime}p)$ data in a wide range of nuclei and in different
kinematics~\cite{book,meucci01,ud93,epja}. 

\section{Inclusive lepton-nucleus scattering}
\label{sec:2}
In the inclusive scattering only the outgoing lepton is detected, the 
final nuclear state is not determined, and all 
elastic and inelastic channels contribute. This requires a different treatment 
of FSI with respect to
the exclusive process, where FSI are described by a complex optical 
potential 
whose imaginary part gives an absorption that reduces the calculated cross
sections. The imaginary part accounts for the fact that in the elastic
scattering, if other channels are open besides the elastic one, part of the
incident flux is lost in the elastically scattered beam and goes to the other
(inelastic) channels which are open. In the exclusive scattering only the
channel $n$ is considered and it is correct to account for the flux lost in the
considered channel. In the inclusive scattering all the channels contribute, 
the flux lost in a channel must be recovered in the other channels, and in the
sum over all the channels the flux can be redistributed but must be conserved.
If the inclusive cross section is obtained from the sum of all the integrated
exclusive one-nucleon knockout processes, due to the interaction of the probe
with all the individual nucleons of the nucleus, and FSI are  described by a
complex optical potential with an absorptive imaginary part, the flux is not
conserved. The use of a complex optical potential seemed inconsistent with the 
requirement of flux conservation and in many calculations based on the 
relativistic IA real potentials have been adopted to describe FSI.

In the simplest approach FSI are simply neglected and the plane-wave
approximation is assumed for the scattering wave functions (RPWIA). In a
different approach an optical potential is used but only its real part is
retained (rROP). The rROP conserves the flux but, independently of its numerical
results, it is conceptually wrong because the optical potential has to be
complex owing to the presence of open inelastic channels. The energy dependence
of the optical potential reflects the different contributions of the inelastic
channels which are open at each energy and, under such conditions, dispersion
relations dictate that the optical potential must have a non vanishing imaginary
part. In a different approach the scattering states are given by the same 
relativistic mean field potential considered in describing the initial nucleon 
states (RMF) \cite{Chiara03,cab1}. The RMF model gives a consistent description of
bound and scattering states, it fulfills the dispersion relations~\cite{hori} 
and also the continuity equation. 

In the Green's function (GF) model  \cite{eenr,ee,cc,eea,eesym,acta,acta1} FSI 
are described  in the inclusive scattering by the same complex optical 
potential as in the exclusive 
scattering, but the imaginary part is used in the two cases in a different way 
and in the inclusive process it redistributes the flux in all the channels
and the total flux is conserved. 

In the GF model, under suitable 
approximations, which are basically related to the IA, the components of the 
hadron tensor are written in terms of the s.p. optical model Green's function. 
This result has been derived by arguments based on the multiple 
scattering theory \cite{hori} or by means of projection operators techniques
within nonrelativistic ~\cite{eenr} and relativistic~\cite{ee,cc,eea} 
frameworks. The explicit calculation of the s.p. Green's function can be 
avoided \cite{eenr,ee,cc}  by its spectral representation, which is based on a 
biorthogonal expansion in terms of a non Hermitian optical-model Hamiltonian 
$\mch$
and of its Hermitian conjugate $\mch^{\dagger}$. 
The components of the hadron tensor are then obtained in the form \cite{ee}
\begin{eqnarray}
 W^{\mu\mu}(q,\omega) = \sum_n \lambda_n \Bigg[ \textrm{Re} T_n^{\mu\mu}
(E_{\textrm{f}}- \varepsilon_n, E_{\textrm{f}}-\varepsilon_n)  \nonumber
\\
- \frac{1}{\pi} \mathcal{P}  \int_M^{\infty} \diff \mathcal{E} 
\frac{1}{E_{\textrm{f}}-\varepsilon_n-\mathcal{E}} 
\textrm{Im} T_n^{\mu\mu}
(\mathcal{E},E_{\textrm{f}}-\varepsilon_n) \Bigg] \ , \label{eq.finale}
\end{eqnarray}
where $E_{\textrm{f}}$ denotes the energy of the final nuclear state,
$\varepsilon_n$ denotes the eigenvalue of the discrete eigenstate $n$ of the 
residual nucleus, and 
\begin{eqnarray}
T_n^{\mu\mu}(\mathcal{E} ,E) &=& \langle \varphi_n
\mid \hat{j}^{\mu\dagger}(\q) \sqrt{1-\mcv'(E)}
\mid\tilde{\chi}_{\mathcal{E}}^{(-)}(E)\rangle \nonumber \\
&\times&  \langle\chi_{\mathcal{E}}^{(-)}(E)\mid  \sqrt{1-\mcv'(E)} \hat{j}^{\mu}
(\q)\mid \varphi_n \rangle  \ . \label{eq.tprac}
\end{eqnarray}
The factor $\sqrt{1-\mcv'(E)}$, where $\mcv'(E)$ is the energy derivative of 
the optical potential, accounts for interference effects between different 
channels and justifies the replacement in the calculations of the Feshbach 
optical potential $\mcv$, for which neither microscopic nor
empirical calculations are available, by the local phenomenological optical 
potential \cite{eenr,ee}. 
Disregarding the square root correction, the second matrix element in 
Eq.~(\ref{eq.tprac}) is the transition amplitude of single-nucleon knockout of 
Eq.~(\ref{eq.dko}), where the imaginary part of the optical potential
accounts for the flux lost in the channel $n$  towards the channels different 
from $n$. In the inclusive response this loss is compensated by a corresponding 
gain of flux due to the flux lost, towards the channel $n$, by the other final 
states asymptotically originated by the channels different from $n$. 
This compensation is performed by the first matrix element in the right hand 
side of Eq.~(\ref{eq.tprac}), that is similar to the matrix element of 
Eq.~(\ref{eq.dko}) but involves the eigenfunction 
$\tilde{\chi}_{\mathcal{E}}^{(-)}(E)$ of the Hermitian conjugate optical
potential, where the imaginary part has an opposite sign and has the 
effect of increasing the strength. As a consequence, in the GF model the 
imaginary part of the optical potential redistributes the flux lost in a 
channel in the other channels, and in the sum over $n$ the total flux is 
conserved.  

The hadron tensor in Eq.~(\ref{eq.finale}) is the sum of two terms. The 
calculation of the second term requires the integration over all the 
eigenfunctions of the continuum spectrum of the optical potential. 
If the imaginary part of the optical potential is neglected, the second term 
in Eq.~(\ref{eq.finale}) vanishes and, but for the square root factor, the 
first term gives the rROP approach.

In the usual applications of the GF model  the matrix elements in 
Eq.~(\ref{eq.finale}) are calculated using the same phenomenological 
bound and scattering states already adopted in DWIA and RDWIA calculations for 
exclusive one-nucleon knockout reactions. For the sum over $n$ a pure shell 
model description is usually assumed: $\varphi_n$ are one-hole states in the 
target nucleus with a unitary spectra strength and the sum is over all the 
occupied states in the shell model. With this simplifying assumption the 
contribution of all the nucleons is correctly included in the inclusive response. 

The GF model allows to recover the contribution of non-elastic channels starting 
from the complex optical potential that describes elastic nucleon-nucleus 
scattering data. 
It provides a consistent treatment of FSI in the exclusive and in 
the inclusive scattering and gives also a good description of $(e,e')$ 
data~\cite{eenr,ee,confee}. 

Both nonrelativistic GF~\cite{eenr,eesym} and relativistic RGF~\cite{ee} 
models have been considered for the inclusive electron scattering, while only
the RGF has been used for CC neutrino scattering \cite{cc}. 
The results of the RMF and RGF models have been compared
in~\cite{confee} for the inclusive electron scattering and in~\cite{confcc} for
the inclusive CC neutrino scattering. 

\begin{figure}
\centerline{
\resizebox{0.98\columnwidth}{!}{%
 \includegraphics{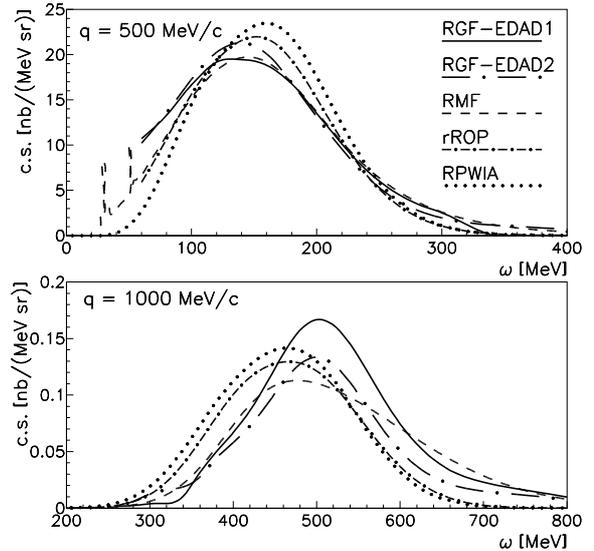} }
}
\caption{Differential cross section of the 
$^{12}$C$(e,e^{\prime})$ reaction for an incident electron energy 
$\varepsilon = 1$ GeV and $q = $ 500 and 1000 MeV$/c$ as a function of the
energy transfer $\omega$. Results for RPWIA (dotted), rROP (dot-dashed),
RGF-EDAD1 (solid),  
RGF-EDAD2 (long dot-dashed), and RMF (dashed) are compared.} 
\label{fig1}   
\end{figure}

An example is displayed in fig.~\ref{fig1}, where the RGF, RMF, rROP, and 
RPWIA cross sections of the $^{12}$C$(e,e')$ reaction calculated in a 
kinematics with a fixed value of the incident electron energy 
($\varepsilon=1$ GeV) and two values of the momentum transfer ($q=500$ and 
1000 MeV$/c$) are compared. Two parameterizations of the ROP have been used 
for the RGF calculations: the energy-dependent and A-dependent EDAD1 (RGF-EDAD1) 
and EDAD2 (RGF-EDAD2) \cite{chc}. 

The RGF and RMF results are always different from the results of the simpler
RPWIA and rROP approaches. 
The differences between RMF and RGF, as well as the differences
between RGF-EDAD1 and RGF-EDAD2, increase with the momentum transfer.  At 
$q$ = 500 MeV$/c$ the three results are similar, both in magnitude and shape, 
at $q = 1000$ MeV$/c$, the shape of the RMF 
cross section shows an asymmetry, with a tail extending towards higher values 
of the energy transfer $\omega$, which is essentially due to the strong energy-independent scalar 
and vector potentials present in the RMF model.
The asymmetry towards higher $\omega$ is less significant but still visible 
for RGF-EDAD1 and RGF-EDAD2, whose cross sections show a
similar shape but a significant difference in magnitude. 
At $q$ = 1000 MeV$/c$ both the RGF-EDAD1 and RGF-EDAD2 cross sections are higher than the 
RMF one in the maximum region, but a stronger enhancement is obtained with 
RGF-EDAD1, which at the peak overshoots the RMF cross section up to 
$40\%$ and it is 
even higher than the RPWIA result.

The differences between the  RGF-EDAD1 and RGF-EDAD2 results are basically due to the 
differences in the imaginary part of the two optical potentials. The 
real terms are very similar for different parametrizations and the rROP cross 
sections are not sensitive to the parameterization considered.

\begin{figure}
\centerline{
\resizebox{0.98\columnwidth}{!}{
 \includegraphics{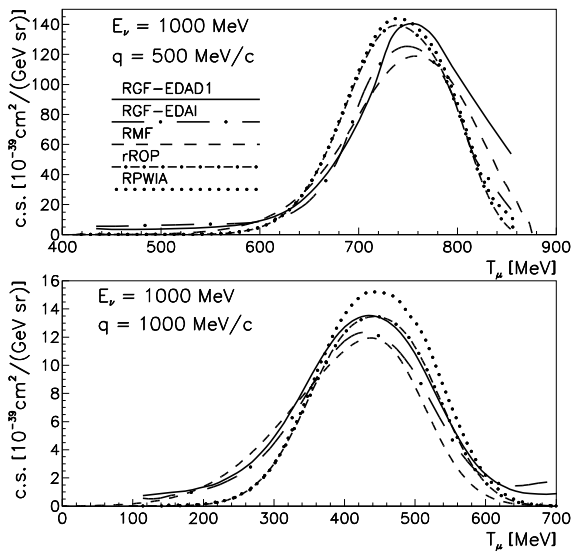} }
}
\caption{Differential cross section of the  
$^{12}$C$(\nu_{\mu} , \mu ^-)$ reaction 
	for E$_{\nu}$ = 1000 MeV and $q$ = 500 MeV/$c$ and 1000 MeV/$c$ as a
	function of the kinetic energy of the outgoing muon $T_\mu$.
 Results for RPWIA (dotted), rROP (dot-dashed) RGF-EDAD1 (solid), RGF-
 EDAI (long dot-dashed), and RMF (dashed) are compared.}
\label{fig2}
\end{figure}

In fig.~\ref{fig2} the RGF, RMF, rROP, and RPWIA results are compared
for the cross sections of the $^{12}$C $(\nu_{\mu},\mu^{-})$ reaction 
calculated with the same incident lepton energy and momentum transfer as for 
the $(e,e^{\prime})$ reaction of fig.~\ref{fig1}. For the RGF model, the RGF-EDAD1 
results are compared  with the 
results obtained with the energy-dependent  but A-independent EDAI potential 
(RGF-EDAI). Also in fig.~\ref{fig2} the shape of the RMF cross section 
shows an asymmetry with a tail extending towards higher values of $\omega$
(corresponding to lower values of the kinetic energy of the outgoing muon
$T_\mu$). An asymmetric shape is shown also by the RGF 
cross sections, while no visible asymmetry is given by the RPWIA and rROP 
results. Also in this case the differences between the two RGF results are due to the different 
imaginary parts of the two ROP's. As already shown for the 
$(e,e^{\prime})$  reaction, the RGF yields in general a larger cross section than the RMF. 

The results in fig.~\ref{fig2} present some differences with respect to the 
corresponding $(e,e^{\prime})$ cross sections in 
fig.~\ref{fig1}. In both cases the differences
between the results of the different models are generally larger for increasing value of 
the momentum transfer. For neutrino scattering, however, such a behavior is 
less evident and clear.  In particular, the RGF-EDAD1 cross section at 
$q$ = 1000 Mev$/c$ does not show the strong enhancement in the region of the 
maximum shown in fig.~\ref{fig1}, where the RGF-EDAD1 result is even larger than 
the RPWIA one. 
In the case of neutrino scattering the RGF results in the region of the 
maximum are generally larger than the RMF ones, but smaller than the RPWIA 
cross sections.
The numerical differences between the RGF results for electron and neutrino 
scattering can mainly be ascribed to the combined effects of the weak current, 
in particular its axial term, and the imaginary part of the ROP \cite{confcc}. 
 
We recall that the RMF model uses as input the real, strong, energy-independent,
relativistic mean field potential that reproduces the saturation properties of 
nuclear matter and of the ground state of 
the nuclei involved. As such, it includes only purely nucleonic contribution 
and does not incorporate any information from scattering reactions. 
In contrast, the RGF uses as input the complex energy-dependent relativistic
optical potential. Phenomenological optical potentials, obtained through a fit 
of elastic proton-nucleus scattering data, are adopted in actual RGF 
calculations. Therefore, the RGF model incorporates information from 
scattering reactions and takes into account not only direct one-nucleon 
emission, but all the allowed final states, as the s.p. Green's function 
contains the full propagator. 

The imaginary part of the ROP includes the overall effect of all the inelastic 
channels, which give different contributions at different energies. This energy dependence makes the RGF results strongly
dependent on kinematics. The differences between the RGF and RMF results can 
be ascribed to the inelastic contributions which are incorporated in the RGF 
but not in the RMF (and  in other models based on the IA), such as, for 
instance, re-scattering processes of the nucleon in its way out of the nucleus, 
non-nucleonic $\Delta$ excitations, which may arise during nucleon propagation, 
as well as to some multinucleon processes. 
These contributions are not included explicitly in the RGF model with a
microscopic approach, but they can be recovered, to some extent, by the 
imaginary part of the phenomenological optical potential. With the use of such 
a phenomenological ingredient, however, we cannot disentangle the role of 
different reaction processes and explain in detail the origin of the
differences, but we can expect that the differences increase with the 
relevance of such inelastic contributions. 

The comparison between the RGF and RMF results can therefore be useful 
to evaluate the relevance of inelastic contributions. 
If in many situations the two models give close predictions (usually
different from those of the simpler RPWIA and rROP), there are also
situations where the differences are large  \cite{confee,confcc}. 

In the comparison with data, we may expect that the RGF can give a better 
description of those experimental cross sections which receive significant
contributions from non-nucleonic excitations and multi-nucleon processes.
This is expected to be the case \cite{Tina10,compmini} of  MiniBooNE data, given the 
nature of the experiment.
While in electron-scattering experiments the beam energy is known and the cross
sections are given as a function of the energy transfer, in neutrino experiments
$q$ and $\omega$ are not known and calculations for the comparison with data 
are carried out over the energy range which is relevant for the neutrino flux. 
The flux-average procedure can include contributions from different kinematic 
regions where the neutrino flux has significant strength and processes other
than direct one-nucleon emission can be important \cite{Ben10,Ben11}. Part of 
these contributions are recovered in the RGF model by the imaginary part of 
the optical potential

\section{Comparison with CCQE MiniBooNE data}
\label{sec:3}
The CCQE $^{12}$C$(\nu_{\mu} , \mu ^-)$ cross sections  
recently measured by the MiniBooNE collaboration \cite{miniboone} have raised 
debate over the role of the theoretical ingredients entering the description of 
the reaction. The experimental cross section is underestimated by the
Relativistic Fermi Gas (RFG) model, and by other more sophisticated models based
on the IA, unless the nucleon axial mass  
$M_A$ is significantly enlarged (1.35 GeV/$c^2$ in the RFG) with respect to the 
accepted world average of all measured values 
(1.03 GeV/$c^2$ \cite{Bern02,bodek}), mostly obtained from deuteron data.
The larger axial mass obtained from the MiniBooNE data on carbon can also be 
interpreted as an effective way to include medium effects which are not taken 
into account by the RFG and by other models.
Before drawing conclusions, it is therefore important to evaluate carefully the 
role played by all the nuclear effects. 

\begin{figure}
\centerline{
\resizebox{0.98\columnwidth}{!}{
 \includegraphics{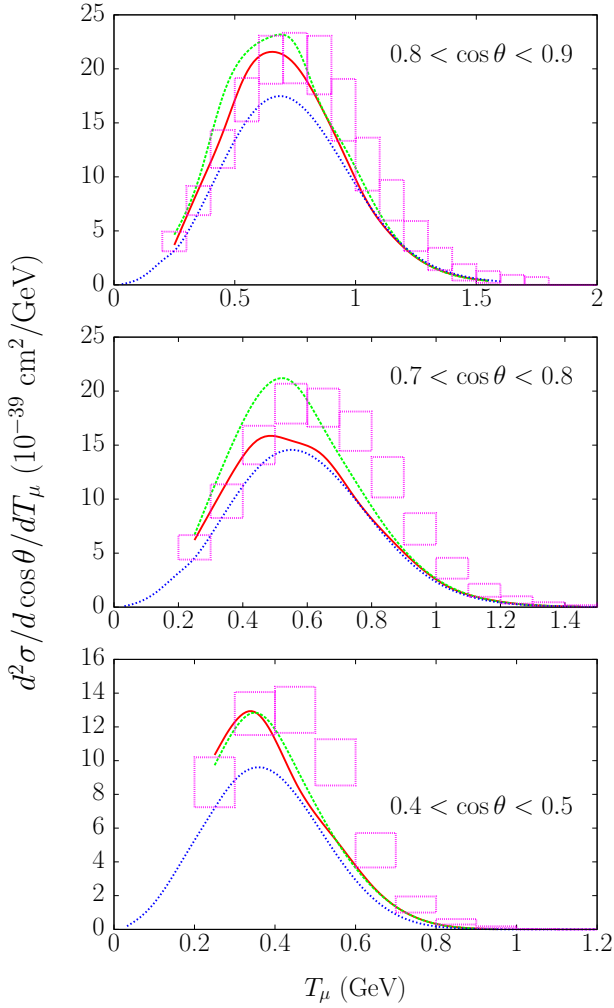} }
}
\caption{Flux-averaged double differential cross section per target nucleon for 
the CCQE $^{12}$C$(\nu_{\mu} , \mu ^-)$ reaction calculated in the RMF 
(blue line) and in the RGF  with EDAD1 (red) and EDAI (green) potentials and 
displayed versus $T_\mu$ for various bins of $\cos\theta$. 
The data are from MiniBooNE~\cite{miniboone}. The uncertainties do not include 
the overall normalization error $\delta N$=10.7\%.}
\label{fig3}
\end{figure}

The effect of FSI is investigated in fig.~\ref{fig3}, where the CCQE 
double-differential $^{12}$C $(\nu_{\mu},\mu^{-})$ cross sections 
averaged over the neutrino flux are displayed as a function of $T_\mu$ for 
various bins of $\cos\theta$, where $\theta$ is the muon scattering angle. 
The RMF results yield reasonable agreement with data for small angles and low 
muon energies, the discrepancy becoming larger as $\theta$ and $T_\mu$ increase.
The shape followed by the RMF and RGF cross sections fits well the slope shown 
by the data. The two models yield close predictions at
larger values of $T_\mu$, for all the bins of $\cos\theta$ shown in 
the figure. The RGF cross sections are generally larger than the RMF ones. 
The differences increase approaching the peak region, where the additional
strength shown by the RGF produces cross sections in reasonable agreement 
with the data. The differences between the RGF-EDAI and RGF-EDAD1 results 
are enhanced in the peak region and are in general 
of the order of the experimental errors.

\begin{figure}
\centerline{
\resizebox{0.98\columnwidth}{!}{
 \includegraphics{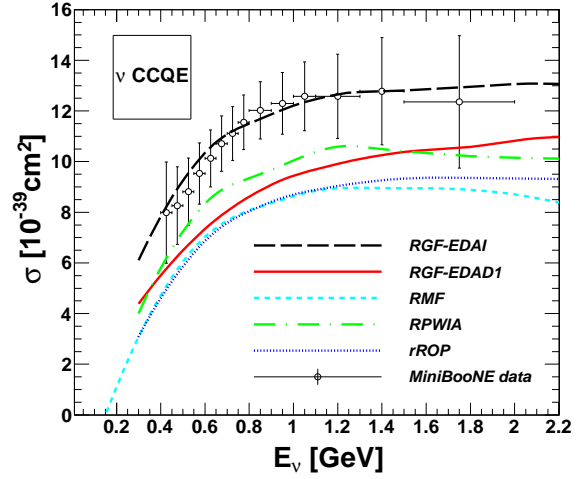} }
}
\caption{Total CCQE cross section per target nucleon as a function of the 
neutrino energy $E_{\nu}$ calculated with the RGF-EDAD1 (solid line), the 
RGF-EDAI (long-dashed line), the RMF (dashed line), the rROP 
(dotted line), and the RPWIA (dot-dashed line). 
The experimental data for neutrino scattering are from MiniBooNE 
\cite{miniboone}}
\label{fig4}
\end{figure}

In fig.~\ref{fig4} the total CCQE cross sections per neutron obtained in 
the RMF, RGF, rROP, and RPWIA models are displayed as a function of the 
neutrino energy 
and compared with the \lq\lq unfolded\rq\rq\ experimental data \cite{miniboone}. 
The rROP, RPWIA, and RMF results usually underpredict the MiniBoone cross section. 
It is shown in \cite{Amaro2011,AmaroAntSusa} that the
differences between models like the RMF, rROP, RPWIA, and superscaling tend to 
be washed out 
in the integration and that all these models, which represent essentially the 
same nucleonic contribution to the inclusive cross sections, undepredict the 
total MiniBooNE CCQE cross section, 
whereas the inclusion of two-particle-two-hole meson-exchange currents  
enhances the results.

Larger cross sections, in particular for larger values of 
$E_\nu$, are obtained in the RGF with both optical potentials. 
The differences between the RGF-EDAI and the 
RGF-EDAD1 results, being RGF-EDAI in good agreement with the shape and 
magnitude of the experimental cross section and RGF-EDAD1 above 
RMF but clearly below the data, are due to the different imaginary parts 
of the two ROP's, particularly for the energies considered in kinematics with 
the lowest $\theta$ and the largest $T_\mu$. 
We notice that EDAI is a single-nucleus parameterization, which does have an 
edge in terms of better reproduction of the elastic proton-$^{12}C$ 
phenomenology \cite{chc} compared to EDAD1, and also leads to CCQE results in 
better 
agreement with data.

The MiniBooNE collaboration has also measured $\bar \nu_{\mu}$ CCQE events.  
The data analysis is currently ongoing \cite{minibooneflux} and some preliminary 
results can be found on the MiniBooNE website \cite{minibooneweb}. 
When available, the antineutrino measurements will be an additional source 
of information about the weak charged-current lepton-nucleus interaction. 

\begin{figure}
\centerline{
\resizebox{0.98\columnwidth}{!}{
 \includegraphics{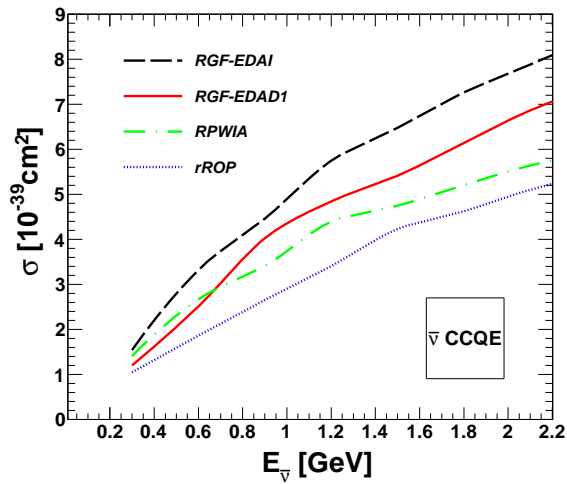} }
}
\caption{Total CCQE cross section per target nucleon as a function of the 
antineutrino energy $E_{\bar\nu}$ 
 calculated with the RGF-EDAD1 (solid line), the RGF-EDAI (dashed line), 
the rROP (dotted lines), and the RPWIA (dot-dashed line). 
}
\label{fig5}
\end{figure}

In fig.~\ref{fig5} the total CCQE cross sections per target nucleon calculated
with the RGF, RPWIA, and rROP for 
antineutrino scattering are displayed as a
function of the antineutrino energy $E_{\bar\nu}$.   
Also for $\bar\nu$ scattering the RGF results are usually larger than the RPWIA 
and rROP ones. The differences between the 
RGF-EDA1 and RGF-EDAI results are significant also 
for antineutrino scattering, 
although somewhat smaller than for neutrino scattering. Moreover, we note that the 
antineutrino cross section does not saturate in the energy range up to 
$\approx$ 2 GeV which we have considered. The different behavior of the cross sections 
calculated for neutrino and antineutrino scattering is related to the relative strength of 
the vector-axial response,  which is constructive in $\nu$ scattering and
destructive in $\bar\nu$ scattering with respect to the longitudinal and 
transverse ones \cite{MeucciAnti}.

The MiniBooNE Collaboration has recently reported \cite{miniboonenc}  also a
measurement of the flux-averaged differential cross section as a function of 
the four-momentum transferred squared, $Q^2 = -q^{\mu}q_{\mu}$, for 
neutral-current elastic (NCE) neutrino scattering on CH$_2$ in a $Q^2$ range up 
to $\approx 1.65\ ($GeV/$c)^2$.
A careful analysis of $\nu$-nucleus NCE reactions introduces additional 
complications, as the final neutrino cannot be measured in practice and a 
final hadron has to be detected: the corresponding cross 
sections are therefore semi-inclusive in the hadronic sector and inclusive in 
the leptonic one. Different relativistic descriptions of FSI are presented and 
compared with the NCE MiniBooNE data in \cite{MeucciNCE}.

\section{Summary and conclusions}
\label{sec:4}

A deep understanding of the reaction mechanism of neutrino-nucleus cross 
sections is mandatory for the determination of neutrino oscillation parameters. 
Reliable theoretical models are required where all nuclear effects are well 
under control. Within the QE kinematics domain, the treatment of FSI is an 
essential ingredient for the comparison with data.  

Models developed for QE electron scattering and successfully tested in 
comparison with electron-scattering data have been extended to neutrino-nucleus 
scattering. 
The results of different relativistic models for the inclusive QE electron and 
neutrino-nucleus scattering, with different descriptions of FSI, have been 
compared in this contribution. In the relativistic plane-wave impulse 
approximation (RPWIA) FSI are simply neglected. In other approaches FSI are 
included in RDWIA calculations where the final nucleon state is evaluated with 
real potentials, either retaining only the real part of the relativistic 
energy-dependent complex optical potential (rROP), or using the same 
relativistic mean field potential considered in describing the initial nucleon 
state (RMF).
In the relativistic Green's function (RGF) model FSI are described in the 
inclusive scattering by the same complex optical potential as in the exclusive 
scattering, but the imaginary part is used in the two cases in a different way 
and in the inclusive process it is responsible for the redistribution of the 
flux in all the channels and the total flux is conserved. 

The differences between the results of the different models depend on
kinematics. The predictions of the RGF and RMF models are close in many 
situations, and generally different from the RPWIA and rROP results. 
There are, however, also situations where the differences are large. 
Larger cross section are generally obtained with the 
RGF model, which is able to give a better description of experimental data. 
In particular, the RGF model is able to reproduce the CCQE MiniBoone cross sections 
without the need to increase the standard value of the nucleon axial mass.
The enhancement of the RGF cross sections is due to the translation to the 
inclusive strength of the overall effect of inelastic channels, due, for 
instance, to re-scattering processes, non-nucleonic excitations, which may 
arise during nucleon propagation, or to some multinucleon processes. Such 
inelastic contribution, which are not incorporated in the RMF and in other 
models based on the impulse approximation, are not included explicitly in the 
RGF model with a microscopic approach, but they can be recovered, to some 
extent, by the imaginary part of the optical potential. The use of 
phenomenological optical potentials, however, does not allow us to disentangle 
the role of different reaction processes and explain in detail the origin of 
the enhancement.

Other models, where multinucleon components are explicitly included, are also 
able to describe the MiniBooNE data without increasing the value of the axial 
mass \cite{Mart1,Mart2,Mart3,Niev1,Niev2}. The important role of contributions 
other than direct one-nucleon emission is therefore confirmed by different and
somewhat alternative models. A careful and consistent evaluation of all 
nuclear effects is required before definite conclusions can be drawn. 
A detailed  comparison of the models and of their results would be helpful for a deeper understanding. 
Processes involving two-body currents, whose role has been discussed in 
\cite{Ben11,bodek,Amaro2011,AmaroAntSusa}, should also be taken into account
explicitly and consistently in a model to clarify the role of multinucleon
emission. Fully relativistic microscopic calculations of two-particle-two-hole 
(2p-2h) contributions are extremely difficult and may be bound to
model-independent assumptions.  

The RGF results are also affected by uncertainties in the determination of
the phenomenological optical potential. 
At present, lacking a phenomenological optical potential which exactly 
fulfills the dispersion relations in the whole energy region of interest, 
the RGF prediction is not univocally determined from the elastic phenomenology.
The differences of the results obtained with different parametrizations
of the relativistic optical potential are produced by the different imaginary 
part, which is the crucial ingredient of RGF calculations. 
It is interesting to notice that the best predictions in comparison with data 
are given by the EDAI potential, that is also able to give the best 
description of the elastic proton-$^{12}$C phenomenology. 
A better determination of a phenomenological relativistic optical potential, 
which closely fulfills the dispersion relations, would be anyhow desirable and 
deserves further investigation.

\section*{Acknowledgements}
We thank F.D. Pacati, F. Capuzzi, J.A. Caballero, J.M. Ud\'{\i}as, and M.B. 
Barbaro for the fruitful collaborations that led to the results reported in 
this contribution.
This work has been partially supported by the Italian MIUR 
through the PRIN 2009 research project.

\end{document}